\documentclass{aa}  
\usepackage{graphicx}
\usepackage{newtxtext}
\usepackage[varg]{newtxmath}
\usepackage{hyperref}
\usepackage{orcidlink}
\usepackage{booktabs}
\usepackage{xcolor}
\usepackage{ulem}

\usepackage{color}
\definecolor{orange}{rgb}{0.9, 0.37, 0.31}
\definecolor{blueish}{rgb}{0.2, 0.2, 0.9}
\definecolor{greenish}{rgb}{0.2, 0.7, 0.2}

\newcommand{\gaia}{\textit{Gaia}\xspace}

\def\bprp{\ensuremath{G_{\rm BP}-G_{\rm RP}}} 
\def\varG{\ensuremath{\mathrm{Var}_G}}

\makeatletter
\DeclareRobustCommand*{\fieldName}[1]{%
  \begingroup\@fieldName\scantokens{\texttt{\small {#1}}\noexpand}\endgroup}
\begingroup\lccode`\~=`\_\relax
   \lowercase{\endgroup\def\@fieldName{\catcode`\_=\active \let~\_}}
\makeatother

\begin{document}

   \title{Cataclysmic Variables and the disc instability model in the \gaia DR3 colour-magnitude diagram}

   \author{Guillaume Dubus\orcidlink{0000-0002-5130-2514}         \and
          Carine Babusiaux\orcidlink{0000-0002-7631-348X}
          }

   \institute{Univ. Grenoble Alpes, CNRS, IPAG, 38000 Grenoble, France\\
              \email{Guillaume.Dubus@univ-grenoble-alpes.fr, Carine.Babusiaux@univ-grenoble-alpes.fr}
        		}

   \date{Received XXX; accepted YYY; in original form \today}

   \titlerunning{CVs and the DIM in the \gaia HRD}
   \authorrunning{Dubus and Babusiaux}
   \abstract
  {Cataclysmic Variables (CVs) are semi-detached binaries composed of a white dwarf orbiting a lower-mass K or M star. }
  {We investigate whether CVs are responsible for a new intriguing feature (the `hook')  that appears in the \gaia DR3 colour-magnitude Hertzsprung-Russell diagram (HRD) when selecting sources with low extinction. We also aim to understand the location of CVs in the HRD based on the predictions of the disc instability model (DIM). The DIM is the foundation on which rests our basic understanding of stable (novae-like) and outbursting CVs (dwarf novae).}
  {We calculate the expected behaviour of CVs in the \gaia HRD taking into account the variable light contributed by the accretion disc, the companion, the white dwarf, and from the bright spot where the Roche lobe overflow stream from the companion intersects the disc.}
  {We find that the `hook' feature is most likely to be composed of CVs. The `hook' corresponds to the limited region where stable CVs (novae-likes) must be located in the HRD according to the DIM, with the bluest systems having the shortest orbital period. Unstable systems, giving rise to dwarf novae outbursts, trace counterclockwise loops in the HRD. The overall behaviour is consistent with the location of the various CV subtypes in the HRD.}
  {These results can be used as a basis to pinpoint interesting outliers in the HRD, either due to their location or their tracks. These outliers may signal new subtypes such as cold, stable CVs with truncated discs, or may challenge the disc instability model.}
  
   \keywords{
                Accretion, accretion discs --
                Novae, cataclysmic variables --
                Stars: dwarf novae --
                Subdwarfs
               }

   \maketitle
%
%-------------------------------------------------------------------

\section{Introduction}

Cataclysmic Variables (CVs) are semidetached binaries composed of a white dwarf orbiting a lower mass K or M star (see \citealt{2001cvs..book.....H} for an introduction, \citealt{2003cvs..book.....W} for a complete monograph, and \cite{2023arXiv230310055W} for a recent review). The companion star fills its Roche lobe, with material lost through the first Lagrangian (L1) point forming an accretion disc around the white dwarf. CVs have orbital periods $P_{\rm orb}$ ranging from $\approx$ 80\,min to 10\,hr, with a gap between 2 and 3\,hr in the distribution (period gap). CVs are thought to be born at the high end of this $P_{\rm orb}$ distribution,  as the companion first fills its Roche lobe and after a common envelope episode has shrunk the orbit of the initial binary. CVs then evolve to shorter $P_{\rm orb}$ as a result of binary angular momentum losses through stellar wind magnetic braking or gravitational wave emission, with the gap due to a drop in the braking efficiency. The population of CVs has thus given insights, constraints, and puzzles to the study of binary stellar evolution.

Accretion in CVs makes for a rich somewhat daunting phenomenology with a long history. A first key is whether the magnetic field of the white dwarf is strong enough to prevent the formation of an accretion disc (polars or AM Her type), to partly truncate the disc (intermediate polars or DQ Her type), or is too weak to influence accretion (non-magnetic CV).  Another key is the variability on timescales of days to years, with a broad separation between novae likes, CVs in a (nearly) permanent luminous state, and dwarf novae, CVs displaying recurrent outbursts with a typical amplitude of 3 to 5 optical magnitudes. Novae likes (NL) subdivide into various types according to their spectrum (RW Tri, UX UMa, SW Sex) or their lightcurve (VY Scl, IW And). Dwarf novae (DN) also subdivide according to their outburst morphology (U Gem, Z Cam, SU UMa, WZ Sge, ER UMa). Finally, novae are CVs where the accreted matter piled up on the white dwarf undergoes runaway fusion, leading to an eruption with a typical amplitude of 8 to 15 magnitudes.  All of these phenomena are essentially related to how accretion proceeds, making CVs a valued proving ground for accretion theory.  

CVs are typically discovered serendipitously in surveys from their blue colour, their spectra with broad emission lines, or their variability. The advent of synoptic spectral (SDSS, LAMOST, etc.) and variability (OGLE, ATLAS, CRTS, ASAS-SN, ZTF, etc.) surveys has vastly increased the number of identified and candidate CVs. The open CV catalogue currently lists more than 14,000 entries \citep{2020RNAAS...4..219J}. \gaia \citep{2016A&A...595A...1G} has been particularly impactful as it removes one of the major uncertainty when studying CVs: their distance.  This has enabled, for example, renewed estimates of the local density of CVs \citep{2020MNRAS.494.3799P,2021MNRAS.504.2420I,2023AJ....165..163C}, constrains on their secular rate of accretion \citep{2022MNRAS.510.6110P}, or a test of the disc instability model for DN and NL \citep{2018A&A...617A..26D}. 

\gaia also provides a treasure trove of information on colours and variability that can be used to distinguish between different classes of objects \citep{2019A&A...623A.110G,2023A&A...677A.137M}. CVs occupy a clearly distinct region in a $\bprp$ colour vs. $G$ magnitude Hertzsprung-Russell diagram (HRD), in between the main sequence and the white dwarf cooling sequence where their lightcurve variability distinguishes them from detached binaries \citep{GaiaDR3vari}.  Moreover, the orbital period of a CV is also related to its location in the HRD \citep{2020MNRAS.492L..40A,2022ApJ...938...46A}. Short orbital period CVs, which have faint companions, are located close to the white dwarf cooling sequence whereas long orbital period systems, harbouring more luminous companions, are located close to the main sequence. Different subtypes of CV also occupy somewhat different locations in the HRD \citep{2018A&A...620A.141R,2019PASJ...71...22I,2020MNRAS.492L..40A,2022ApJ...938...46A}. 

Previous \gaia studies dedicated to CVs used compilations of known objects and did not correct for extinction. Here, we investigate a sample of \gaia sources selected for low extinction (Section 2). Their average position in the HRD shows a new intriguing feature that we attribute to CVs. To understand this feature, as well as the tracks left by individual CVs in the HRD as a result of their variability, we calculate the expected behaviour of CVs in the HRD taking into account the variable light contributed by the accretion disc. To do this, we model the accretion disc as a thin, radiatively efficient, $\alpha$ disc. Such discs are stable at high mass accretion rates (NL) but become unstable if their temperature becomes low enough that ionised hydrogen recombines (DN). The large jump in opacity triggers a thermal-viscous instability and cycles of outbursts. This disc instability model (DIM) forms the basis to understand CV lightcurves \citep[see ][for reviews]{Lasota:2001th,2020AdSpR..66.1004H}. We examine the HRD track predicted by the DIM for CVs in section 3. We then discuss the consistency of CV locations in the HRD diagram with the DIM (section 4) before concluding.

\section{The CV area seen by \gaia}

\begin{figure*}
    \centering
    \includegraphics[width=0.245\textwidth]{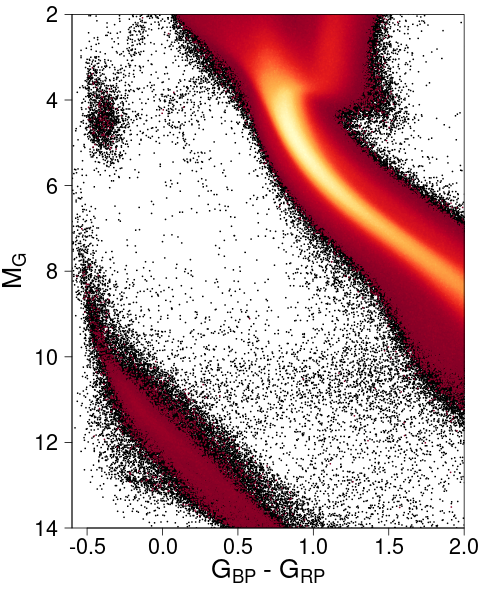}
    \includegraphics[width=0.245\textwidth]{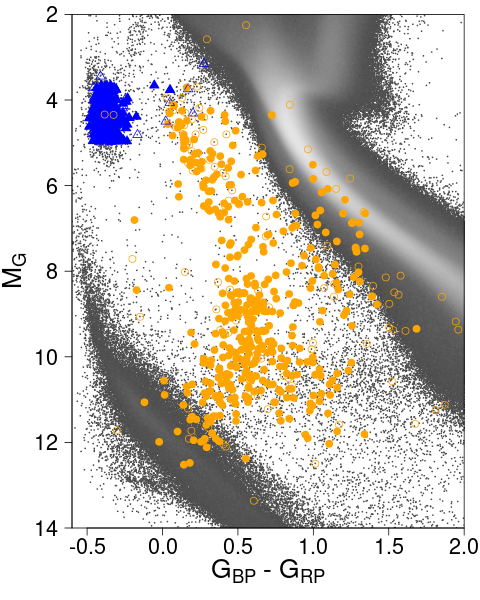}
    \includegraphics[width=0.245\textwidth]{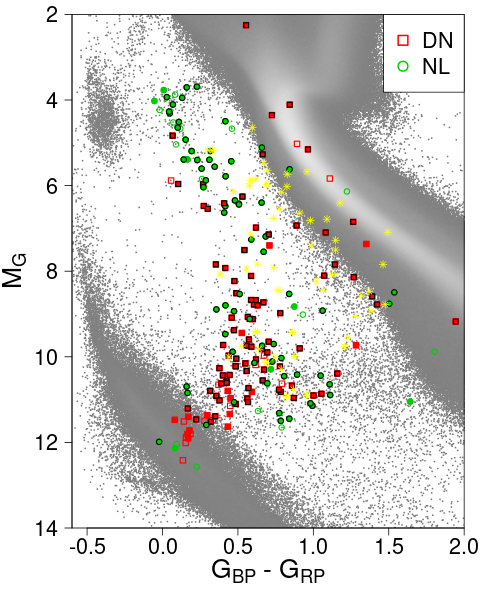}
    \includegraphics[width=0.245\textwidth]{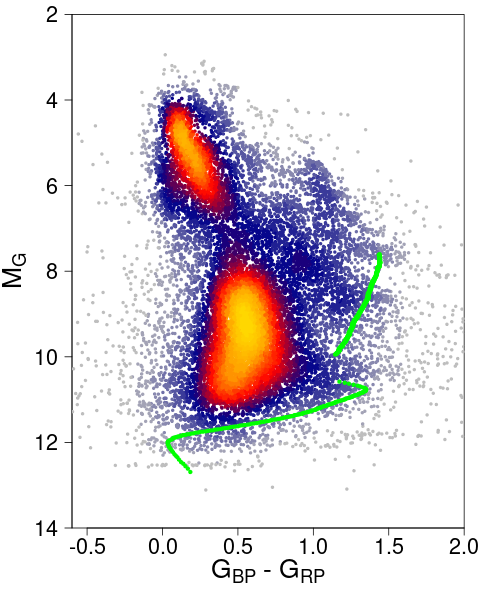}
     \caption{
     {\it Left}: \gaia DR3 low extinction HRD ($A_0<0.1$~mag, parallax uncertainty smaller than 10\%, ruwe$<$1.4, $C^*<10 \times \sigma_C^*$).
     {\it Middle left:} HRD overlaid with \gaia DR3 variability types 'CV' (orange circles) and 'sdB' (blue triangles). Filled symbols correspond to variability score (\fieldName{best_class_score}) $>0.5$.
     {\it Middle right:} CV from the \cite{Ritter03} catalogue with $A_0<0.1$ and parallax uncertainty smaller than 10\%. Filled symbols correspond to sources with \gaia DR3 variability detected and symbols with black borders are sources classified as 'CV' in \gaia DR3. Yellow stars corresponds to CV removed by the ruwe$<$1.4 or $C^*<10 \times \sigma_C^*$ criteria. 
     {\it Right:} heatscatter plot of all the epoch photometry data points of the CV variables with \fieldName{best_class_score}$>0.5$ (all filled orange circles in middle left panel). The \cite{2011ApJS..194...28K} white-dwarf + main-sequence star tracks are shown in green.}
    \label{fig:hrds}
\end{figure*}

\begin{figure*}
\begin{minipage}[c]{\columnwidth}
    \centering
    \includegraphics[width=0.8\columnwidth]{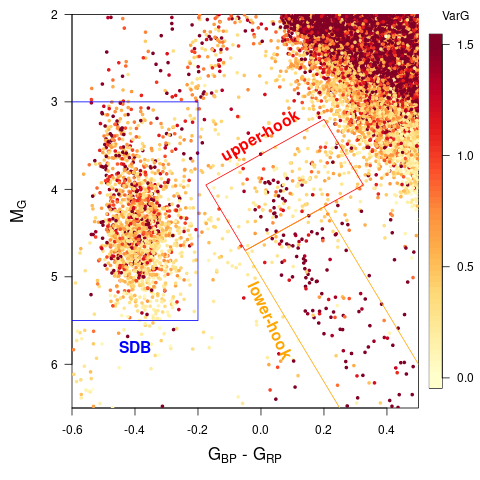}
     \caption{HRD colour-coded by the mean \varG\ of all stars with $A_0<0.1$, parallax uncertainty smaller than 10\%, ruwe$<$1.4 and $C^* < 10 \times \sigma_{C^*}$. The colour scale saturates for values lower than 0 and higher than 1.5. Boxes corresponds to areas discussed in the text: the sdB (blue), the upper hook (red) and the lower hook (orange).
     }
    \label{fig:varG}
\end{minipage}
\hfill
% binarity
\begin{minipage}[c]{\columnwidth}
    \centering
    \includegraphics[width=0.8\columnwidth]{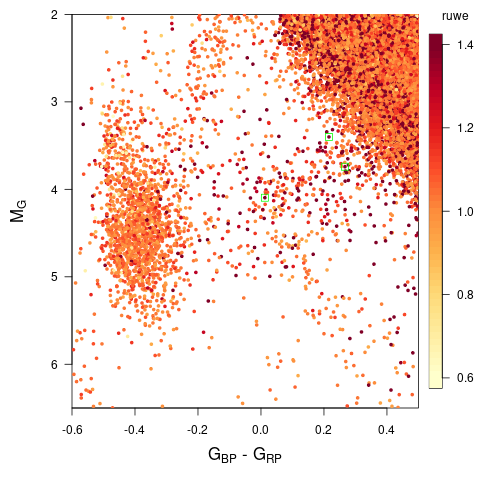}
     \caption{HRD colour-coded by the \fieldName{ruwe} of all stars with $A_0<0.1$, parallax uncertainty smaller than 10\%, $C^* < 10 \times \sigma_{C^*}$, \fieldName{ipd_frac_multi_peak<3}. The colour scale saturates for values lower than 0.6 and higher than 1.4. Green squares correspond to orbital solutions provided by \gaia DR3. 
     }
    \label{fig:ruwe}
\end{minipage}
\end{figure*}

Figure~\ref{fig:hrds} (left panel) shows a zoom of the update of the \gaia low-extinction HRD \citep[Fig.~5 of][]{GaiaDR2hrd} updated to the \citet{GaiaEDR3} EDR3 data and \citet{2022A&A...661A.147L} extinction map. We selected stars with $A_0<0.1$~mag, relative parallax error better than 10\%, relative flux error better than 50\% in $G$ and 20\% in $G_{BP}$ and $G_{RP}$, $\fieldName{ruwe}<1.4$\footnote{\fieldName{ruwe} is the astrometric renormalised unit weight error which is $\sim1$ for astrometrically well-behaved sources \citep{2021A&A...649A...2L}. The criteria $\fieldName{ruwe}>1.4$ is typically used to select problematic astrometric solutions such as astrometric binaries \citep{GaiaDR3nss}.}, and a corrected BP and RP flux excess $C^*$ lower than $10 \times \sigma_{C^*}$ \citep{GaiaEDR3phot}\footnote{The total flux measured in BP and RP should be close to the G-band flux so that $C=(F_{BP}+F_{RP})/F_G$ should be close to one. $C^*$ takes into account the colour dependency of $C$ and is expected to be close to zero, with a scatter $\sigma_{C^*}$ that depends on the $G$ magnitude and with large values expected in case of contamination of the BP or RP spectra by other nearby sources \citep{GaiaEDR3phot}.}. 
This selection leads to more than 26 millions stars, 6 times more than used in \cite{GaiaDR2hrd}. This HRD shows a new prominent feature at $M_G\sim4$ and $\bprp\sim0.2$ going from the main sequence to the hot subdwarf (here after called sdB) clump that seems to turn at $\bprp\sim0$ and continue down, parallel to the main sequence. We call this feature the `hook', and we distinguish two areas for the discussion hereafter: the `upper' hook going from main sequence to sdB clump and the `lower' hook going parallel to the main sequence (see Fig.~\ref{fig:varG}).

The hook region is in the area of the HRD where we expect to see systems with white dwarfs or hot-subdwarfs plus main-sequence or giant companions. 
There are 273 sources in the upper hook and 304 sources in the lower hook, of which 9\% and 34\% (respectively) have been associated to a DR3 variability type in \gaia DR3 \citep{2023A&A...674A...1G, GaiaDR3vari}. Besides the wide-ranging short-timescale variability type, which flags systems with variability on a $<0.5-1$\,day timescale, we find that only two variability types populate the hook: cataclysmic variables (`CV') and subdwarf B stars (`sdB'). These are represented in Fig.~\ref{fig:hrds} (middle left panel), selected with the same criteria as those used to create the main \gaia DR3 HRD (except for the relative flux error criteria, which is removed to allow large relative flux error due to variability in CVs).  
Both CVs and sdBs are found on the upper hook, while only the CVs follow the full trend going down parallel to the main sequence. The right panel of Fig.~\ref{fig:hrds} shows the heatscatter plot of the \gaia-identified CVs. There are two clumps: one along the lower hook and another close to the white dwarf sequence, consistent with the overall CV distribution. The heatscatter plot is clearly bounded at low luminosity by the white dwarf plus companion tracks taken from \citet[][]{2011ApJS..194...28K} (see Appendix~\ref{sec:GaiaMags}). The measurements are brighter and bluer as expected since the accretion disk and bright spot also contribute to the overall light. We note that the shortest period tracks of \citet{2011ApJS..194...28K} correspond to the peak of \gaia DR3 variables classified as white dwarfs (`WD'), which are indeed known to be contaminated by white dwarfs with a cooler companion causing large variations in $G_{BP}$ and $G_{RP}$ \citep{Rimoldini23}.

To further characterise the \gaia-identified CV population, we compare in Fig.~\ref{fig:hrds} (middle right panel) with CVs taken from the \cite{Ritter03} catalogue using the same quality criteria ($A_0<0.1$~mag, relative parallax error better than 10\%,  $\fieldName{ruwe}<1.4$, $C^* < 10 \times \sigma_{C^*}$). We show only CVs of DN or NL type, merging SU UMa with DN and VY Scl with NL for clarity. This figure clearly indicates that CVs located on the hook are novae-like. The quality criterion on \fieldName{ruwe} tends to remove possible astrometric binaries, while the criterion on the flux excess factor tends to remove polluted spectra. Their overall impact is to remove sources whose location in the HRD is contaminated by a nearby unrelated star or by an unknown tertiary companion, making the CV appear redder than it is. Indeed, the yellow stars (middle right panel, Fig.~\ref{fig:hrds}) show that these criteria remove mainly CVs from \cite{Ritter03} that are located redward of the hook. 

% variability

While the blue part of the upper hook seems  to be the continuation of the lower hook and is clearly associated to CVs according to Fig.~\ref{fig:hrds}, the origin of the red part of it is less obvious. In the following we compare the properties of the upper and lower parts of the hook.
The upper hook region of HRD is known to contain wide binary systems \citep[see e.g. Fig.~12 of][]{Culpan22}. Figure~\ref{fig:ruwe} shows the distribution of the renormalised unit weight error of \gaia astrometry (\fieldName{ruwe}). Following \cite{GaiaDR3nss}, we select stars with \fieldName{ipd_frac_multi_peak<3}\footnote{\fieldName{ipd_frac_multi_peak} is the fraction of observations where more than one peak is detected during the image parameter determination (IPD).} to avoid visual double stars. The upper hook is found to have larger \fieldName{ruwe} than the lower hook area (confirmed by a K-S test with p-value 3e-10). Within the Fig.~\ref{fig:ruwe} sample, we have 3 stars with a \gaia DR3 NSS solution \citep{GaiaDR3nss}, two astrometric and one spectroscopic. They have periods between 617 and 828 days. Part of the sources in the upper hook are likely to be sdBs in binaries \citep{Culpan22}. If sources with large \fieldName{ruwe} are CVs then they are likely to be triples as \gaia is more sensitive to longer orbital periods than the typical CV orbital period. For example, \object{TX Col}, a NL CV located on the hook but not included in our selection, has a wide component resolved by \gaia at 2.5 arcsec \citep{2020MNRAS.494.1448I}.

 % colour
\begin{figure}
    \centering
    \includegraphics[width=0.8\columnwidth]{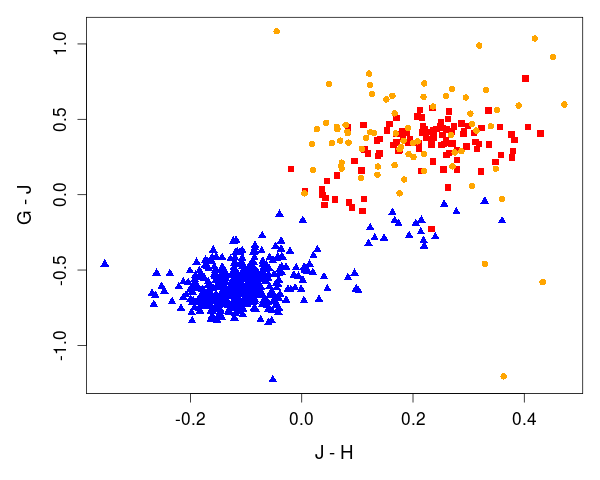}
     \caption{Colour-colour diagram of stars in the areas indicated in Fig.~\ref{fig:varG}, selected with $A_0<0.1$, parallax uncertainty smaller than 10\%, \fieldName{ruwe}$<1.4$, $C^* < 10 \times \sigma_{C^*}$ and 2MASS Qflag `AAA'. Blue triangles correspond to stars in the SDB area, red square in the upper-hook area and orange circles in the low-hook area.
     }
    \label{fig:CC2mass}
\end{figure}

Using a colour-colour diagram does not help to distinguish sdBs from CVs in the hook. Figure~\ref{fig:CC2mass} reproduces Fig.~5 of \cite{Green08} who showed that the `pure' spectrum sdBs (bottom left of Fig.~\ref{fig:CC2mass}) are well-separated from the composite spectrum sdBs. However, there is no clear difference between composite sdBs and CVs in this diagram: both upper hook and lower hook stars have similar colours and fall in the composite spectrum sdB area. 

Kinematics can help. We compared the tangential velocity of the three areas and found that the sdBs and the upper hook share a close distribution (K-S test p-value=0.01) which contains thick disc and halo kinematics \citep[see also][]{Saffer01} while the lower hook has kinematics closer to the thin disc only. Again, this is consistent with the upper hook containing both sdBs and CVs, while the lower hook contains only CVs.\footnote{We also tested for differences in the kinematics of the novae-like and dwarf novae plotted in Fig.~\ref{fig:hrds}, using either the \cite{Ritter03} classification or the \gaia CV classification with a rough separation at $M_G=7$, finding that there is none.}

To look at the variability properties of the hook, we use the \varG index \citep{BarberMann23}: 
\begin{equation}
    \varG = \log_{10}\left(\frac{\sigma_F}{<F>}\right) - \log_{10}\left(\sigma_p(G,N_{\rm obs})\right)
\end{equation}
with $\sigma_p(G,N_{\rm obs})$ the fitted \gaia photometric uncertainty obtained from the \cite{GaiaEDR3phot} tool\footnote{\url{https://github.com/gaia-dpci/gaia-dr3-photometric-uncertainties/}} and with $\frac{\sigma_F}{<F>}$ the inverse of the \fieldName{phot_g_mean_flux_over_error} parameter. Figure~\ref{fig:varG} shows that the right part of the upper hook corresponds to an area without specific photometric variation opposite to the lower hook which shows a strong photometric variation.  This is consistent with the heatscatter plot of CVs (right panel, Fig.~\ref{fig:hrds}). There are also strongly variable systems in the sdB clump, with a fraction of \gaia DR3 variable of 31\%, similar to the lower hook area. 

\begin{figure}
    \centering
    \includegraphics[width=0.8\columnwidth]{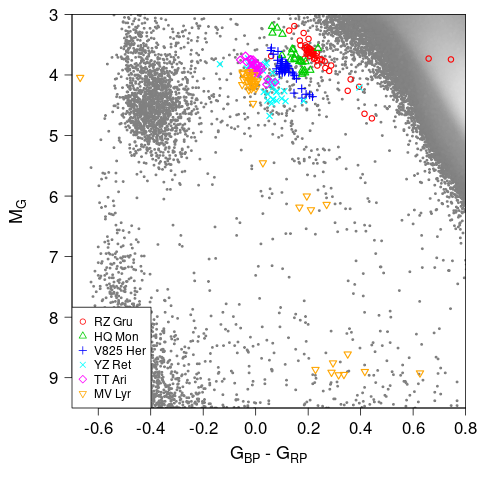}
     \caption{Variation in the HRD of the intrinsically brightest \gaia DR3 CV of Fig.~\ref{fig:hrds} located in the upper-hook area.}
    \label{fig:brightestnovae}
\end{figure}
Figure~\ref{fig:brightestnovae} shows the variability in the HRD of the six brightest (in absolute magnitude) \gaia DR3 CV of the upper-hook with $A_0<0.1$ and parallax uncertainty smaller than 10\% (\object{RZ Gru}, \object{HQ Mon}, \object{V825 Her}, \object{YZ Ret}, \object{TT Ari}, \object{MV Lyr}). All are novae-like systems. MV~Lyr is a VY~Scl NL that went down to its low state during the \gaia observation time. YZ~Ret erupted  as a classical novae after the time period covered by DR3 leading to the \gaia Science Alert Gaia~20elz. 
The HRD variability of these CVs distinguishes them from sdBs. Indeed, we find that the variables classified as sdB by \gaia and located in the upper-hook area all have horizontal variability in the HRD (see Fig.~\ref{fig:hookS} for an example), similarly to the sdBs in the main clump and unlike the CVs shown in Fig.~\ref{fig:brightestnovae}.

\section{Expectations from CV models}
Here, we examine what the disc instability model has to say about the location of CVs in the \gaia HRD. The location of a given CV in the HRD is set by combining the light from its stellar components, the accretion disc and its bright spot. The companion fills its Roche lobe so its mass $M_2$ and radius $R_2$ are essentially set by the system orbital period $P_{\rm orb}$ once the white dwarf mass $M_1$ is fixed \citep{2011ApJS..194...28K}. However, departures away from the main sequence are expected, especially at longer $P_{\rm orb}$ where the companion must be evolved to fill its Roche lobe. In the following, we set $M_1=0.75\rm\,M_\odot$ as in \citet{2011ApJS..194...28K}. We investigated CVs with $P_{\rm orb}$ between 80\,min and 7\,hr, resulting in companions with masses ranging from $\approx 0.07\rm\,M_\odot$ (period minimum) to $\approx 0.75\rm\,M_\odot$ ($P_{\rm orb}=$7\,hr). We refer to Appendix~\ref{sec:GaiaMags} for details on the assumptions made to compute the accretion disc, bright spot, star, and white dwarf contributions to the total emission from the system. We do not consider the impact of interstellar extinction on the model magnitudes since we compare to sources selected for $A_0<0.1$~mag (Section 2).

The accretion parameters can be set once the stellar parameters are known. The contribution of the bright spot, where the material falling from the L1 point shocks the material in the outer disc, depends on the assumed mass transfer rate from the companion $\dot{M}_{\rm t}$. The accretion disc contribution, modelled as a multi-colour blackbody, depends on the radial distribution of temperature. We start with expectations from steady disc before considering more complex assumptions.

\subsection{Steady disc}
For a steady disc, the mass accretion rate $\dot{M}$ is constant throughout the disc and equal to the mass transfer (or mass loss) rate $\dot{M}_{\rm t}$ from the companion to the accretion disc through the L1 point. The radial distribution of the effective temperature $T_{\rm eff}$ as a function of disc radius $R$ is given by \citep{Shakura:1973vo}
\begin{equation}
\sigma T_{\rm eff}^4 = \frac{3 GM\dot{M}}{8\pi R^3}\left[ 1-\left(\frac{R}{R_{\rm in}}\right)^{1/2}\right]
\label{eq:teff}
\end{equation}
with $R_{\rm in}$ the radius of the inner edge of the disc. We start by assuming that the accretion disc extends all the way to the white dwarf, hence setting $R_{\rm in}$ equal to the radius of the white dwarf $R_1$. The outer disc radius $R_{\rm disc}$  is a function of the mass ratio $q=M_2/M_1$ and the orbital separation $a$. We set $R_{\rm disc}$ from the average of the tabulated values of $R_{\rm disc}/a$ in \citet{1975ApJ...198..383L} and \citet{1977ApJ...216..822P}. 

\begin{figure*}
    \centering
    \includegraphics[width=0.245\textwidth]{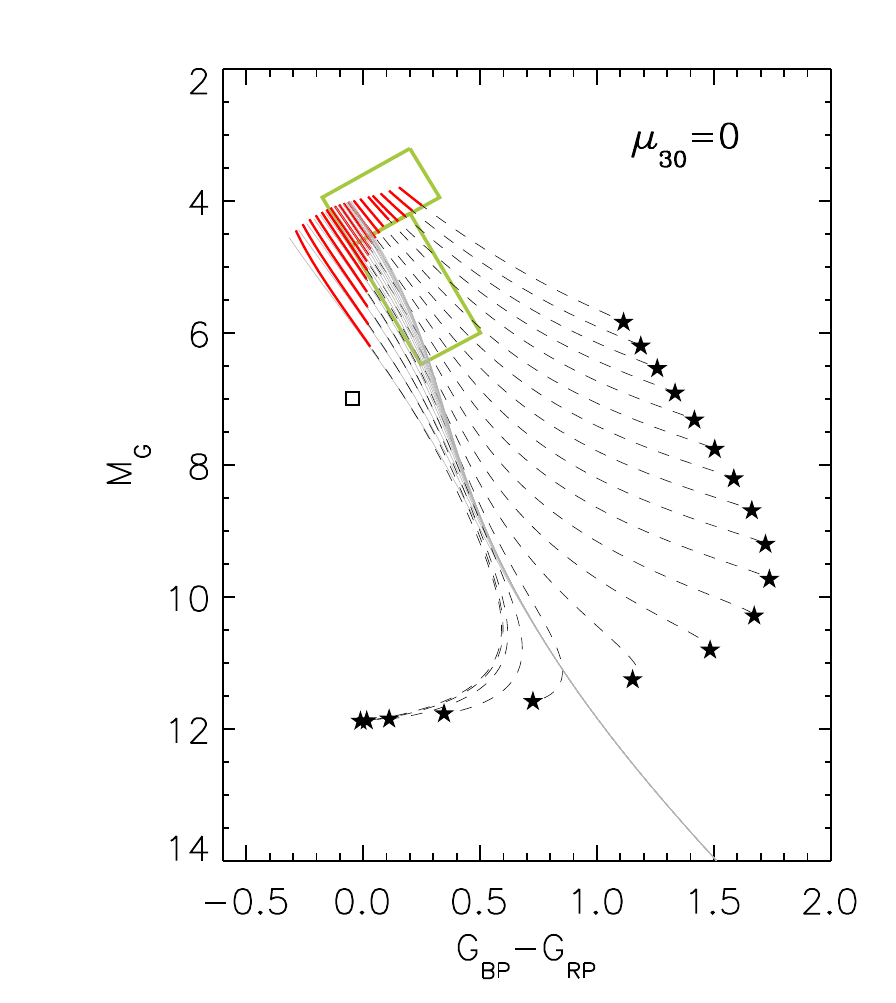}
    \includegraphics[width=0.245\textwidth]{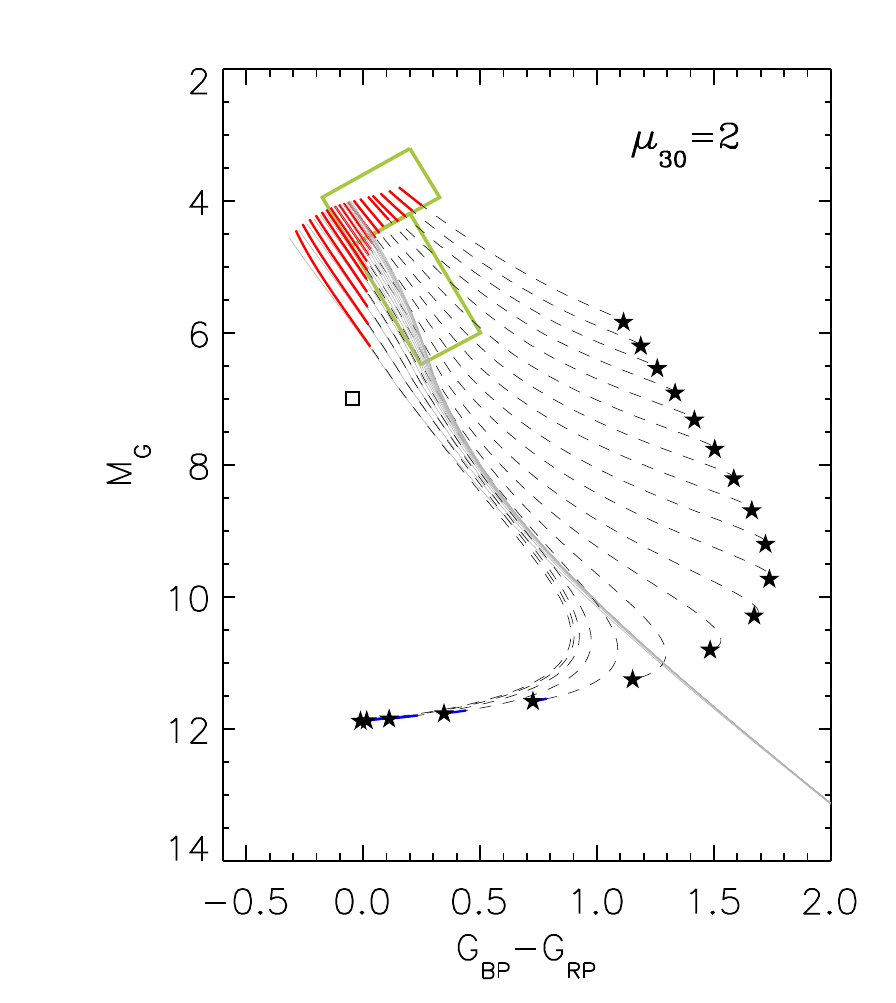}
    \includegraphics[width=0.245\textwidth]{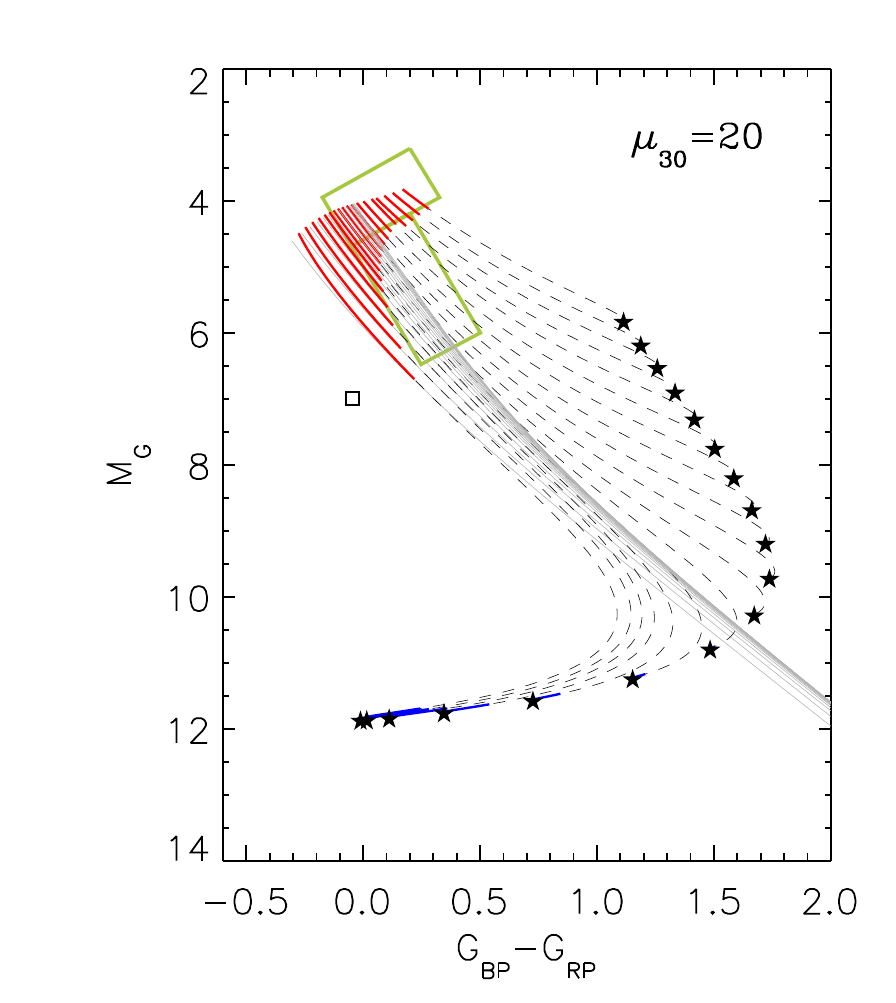}
    \includegraphics[width=0.245\textwidth]{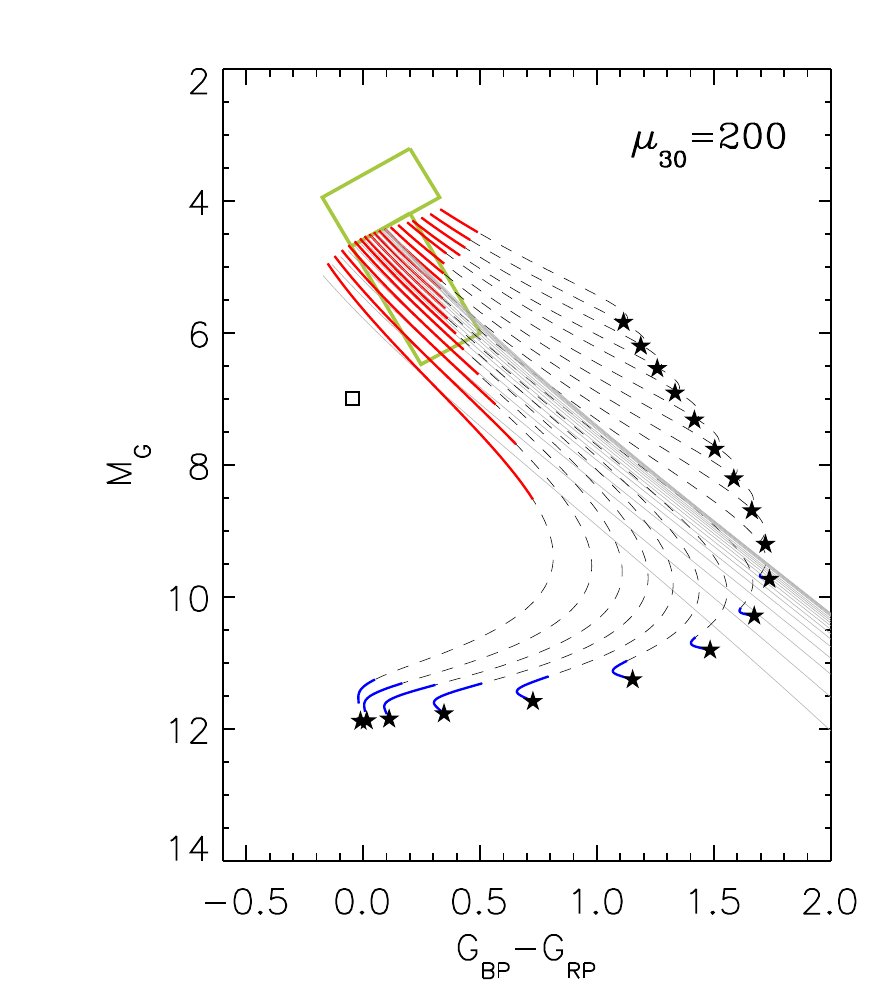}
     \caption{Sequences of colour magnitude positions assuming a steady accretion disc and, for each panel, a given value of the white dwarf dipole moment ($\mu_{30}=\mu/10^{30}\rm\,G\,cm^3$). Each sequence is obtained by varying $\dot{M}$ up to $10^{18}\rm\,g\,s^{-1}$ for a given $P_{\rm orb}$. The orbital period increases from the bottom left sequence ($P_{\rm orb}=80\rm min$) to the top right sequence ($P_{\rm orb}=7\rm\,hr$). The thin grey lines trace only the contribution of the disc. The dashed lines add light from the companion star, the white dwarf, and the bright spot. The stars indicate the location of the combined light from the companion and white dwarf for each $P_{\rm orb}$ sequence. The square indicates the position of the most luminous bright spot emission. When $\mu_{30}\neq 0$, the inner disc is truncated at the magnetospheric radius $R_{\rm m}$. The thick red (resp. blue) lines overplotted on the dashed lines highlight the location of hot (resp. cold) stable steady discs. The green rectangles delimit the upper and lower hook regions defined in Fig.~\ref{fig:varG}.}
    \label{fig:steady}
\end{figure*}

Figure \ref{fig:steady} shows where the systems are located in the HRD with these assumptions. The lines show sequences obtained by varying the disc mass accretion rate $\dot{M}$ from $10^{12}$ to $10^{18}\rm\,g\,s^{-1}$ for a given $P_{\rm orb}$. We plot 18 sequences, corresponding to $P_{\rm orb}$ varying from 80\,min to 7\,hr in steps of 20\,min. The thin continuous grey lines trace only the accretion disc contribution in the HRD, whereas the dashed lines also include the stellar and bright spot contributions. As expected, the disc is bluer and more luminous as $\dot{M}$ increases (Eq.~\ref{eq:teff}). discs are also more luminous for long $P_{\rm orb}$ as they grow larger. However, at low $\dot{M}$ the emission is a sum of Rayleigh-Jeans contributions and all tracks converge in the HRD.

Comparing the dashed lines with the thin grey lines shows that the contributions from the other sources of light in the system cannot be ignored to position systems in the HRD. The stars trace the location of the combined light from the white dwarf and companion star. The contribution from these sources of light was taken into account following the assumptions set out in Appendix~\ref{sec:GaiaMags}. At low $P_{\rm orb}$, light from the donor star makes a negligible contribution and the star symbols join the white dwarf sequence. Inversely, the stars join the stellar main sequence in the HRD at high $P_{\rm orb}$ as the white dwarf contribution becomes negligible. At low $\dot{M}$ the contribution of the accretion disc becomes negligible so that the dashed lines connect to the star symbols. At high $\dot{M}$ the accretion disc dominates, but the companion star significantly reddens the colour at long orbital periods. Another source of light is the bright spot emission, which varies with $\dot{M}$ and $P_{\rm orb}$ (see Appendix~\ref{sec:GaiaMags} for details). Its maximum value, shown as a square, is reached for $P_{\rm orb}=80\rm\,mn$ and $\dot{M}=10^{18}\rm\,g\,s^{-1}$. It has little influence on the HRD sequences plotted here, except at the shortest orbital periods where it slightly reddens the colour at maximum $\dot{M}$ compared to disc-only emission.

The vertical thermal equilibrium of a disc annulus with a temperature  $\approx 6500\rm\,K$ becomes unstable due to the sharp change in opacity associated with hydrogen ionisation. The instability leads to an  outburst phase with a high accretion rate followed by quiescence during which the accretion rate through the disc is low \citep{Lasota:2001th}. In order to be hot and stable, with hydrogen ionised everywhere, the mass accretion rate through the disc must be
\begin{equation}
\dot{M}\ge \dot{M}_{\rm hot}=8.1\times 10^{15}\, \left( \frac{R_{\rm disc}}{\tiny 10^{10}\rm\,cm}\right)^{2.64} \left(\frac{M_{1}}{\tiny 1\rm\, M_{\odot}}\right)^{-0.89}\rm\,g\,s^{-1}
\end{equation}
using the criterion for Solar composition discs derived in \citet{2008A&A...486..523L}. Similarly, the disc is cold and stable, with hydrogen recombined everywhere, if the mass accretion rate is below 
\begin{equation}
\dot{M}\le \dot{M}_{\rm cold}=6.9\times 10^{12}\, \left( \frac{R_{\rm in}}{\tiny 10^{9}\rm\, cm}\right)^{2.58}\left(\frac{M_{1}}{\tiny 1\rm\, M_{\odot}}\right)^{-0.89}\rm\,g\,s^{-1}
\label{eq:cold}
\end{equation}
The thick red lines in Fig.~\ref{fig:steady} highlight the portion of the dashed lines where the criteria for a hot and stable disc is realised. Hot, steady discs (novae-like) occupy only a small region of the HRD. Most of the sequences in the HRD are actually unstable and would lead to outburst cycles (dwarf novae). Cold and stable discs are also possible at very low $\dot{M}$, where the contribution of the disc is negligible. These are indistinguishable from the star symbols in this figure.

\subsection{Truncated discs}
White dwarfs can have a magnetic field that is strong enough to truncate the inner disc (intermediate polars). Figure \ref{fig:steady} also shows how the colour-magnitude sequences change when the inner disc is truncated at the magnetospheric radius
\begin{equation}
R_{\rm m}=6.2\times 10^{8} \left(\frac{\mu}{\tiny 10^{30}\rm\, G\,cm^{3}}\right)^{4/7} \left(\frac{M_{1}}{\tiny 1\rm\, M_{\odot}}\right)^{-1/7}\left(\frac{\dot{M}}{\tiny 10^{16}\rm\, g\,s^{-1}}\right)^{-2/7}\rm\,cm
\end{equation}
where $\mu=BR_{1}^{3}$ is the dipole moment and $B$ is the magnetic field at the white dwarf surface. The magnetospheric radius can also be a good proxy for the radius at which an $\alpha$ disc transitions to magnetic wind-dominated accretion, when the torque exerted by a large-scale magnetic field takes over the transport of angular momentum \citep{2019A&A...626A.116S}.

The tracks do not all join at low $\dot{M}$ when a truncated disc is considered (thin grey lines in Fig.~\ref{fig:steady}). This is because $R_{\rm in}=R_{\rm m}$ is now also changing. Increasing $\mu$ truncates the inner disc increasingly further out. Stable cold discs are now possible at a higher $\dot{M}$ (Eq.,\ref{eq:cold}). They appear as thick blue lines at high $\mu$ in Fig.~\ref{fig:steady}. Additionally, the thick red lines materialising where a hot stable disc is possible cover a larger fraction of the HRD when $\mu$ is large: the truncation lowers the overall magnitude but also significantly reddens the colour when the high temperature inner region is removed. The disc can become entirely truncated for large $\mu$ and accretion occurs along magnetic field lines (polar CVs).

\subsection{Erupting discs}

\begin{figure}
    \centering
    \includegraphics[width=0.33\textwidth]{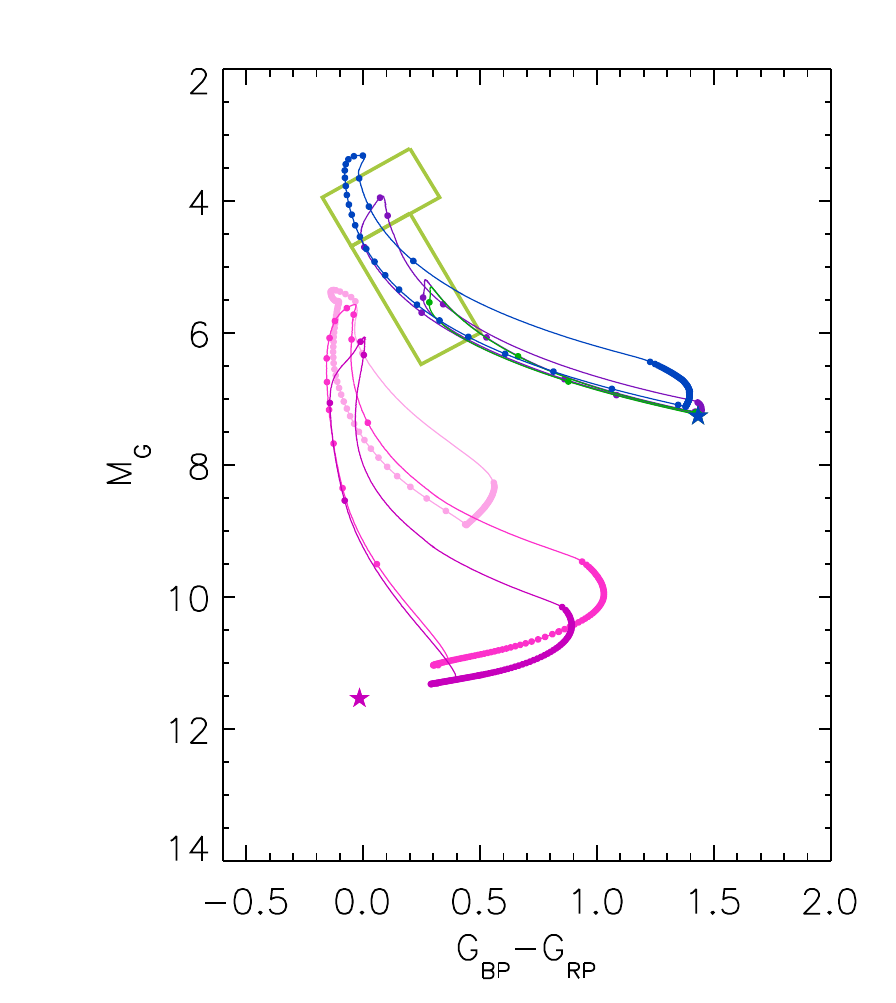}
     \caption{Colour-magnitude tracks traced in the HDR by representative CVs with unstable accretion discs. Two sets of tracks are plotted, corresponding to $P_{\rm orb}=88\rm\,min$ (bottom set, pinkish colours) and $P_{\rm orb}=6\rm\,hr$ (top set, bluish colours). Three tracks are plotted in each set, corresponding to 3 different mass transfer rates $\dot{M}_{\rm t}$. The dots on each track regularly sample in time the outburst cycle. The lightcurves are shown in Fig.~\ref{fig:lightc}. The green rectangles delimit the upper and lower hook regions defined in Fig.~\ref{fig:varG}.}
    \label{fig:loop}
\end{figure}

Figure~\ref{fig:loop} shows examples of colour-magnitude tracks followed by CVs with unstable discs (dwarf novae). Heating and cooling fronts propagate through the disc when it is unstable, cycling the disc from a hot, high accretion outburst state to a cold, low accretion quiescent state and back again (see the review by \citealt{Lasota:2001th}). The outburst cycles depend on $M_{1}$, $\dot{M}$, and on the $\alpha$ parameter for turbulent angular momentum transport \cite{Shakura:1973vo}. The value of $\alpha$ must be taken higher in the hot state ($\alpha_{\rm h}$) than in the cold state ($\alpha_{\rm c}$) to reproduce CV outbursts. The outburst cycles also depend on the variations of the inner and outer disc radii during the cycle. The disc is expected to grow during outburst, as increased accretion requires increased angular momentum transported outwards. The radial distribution of the temperature varies strongly during the cycle. It is close to a steady disc at the peak of the outburst and flattish during quiescence. The time evolution of these radial distributions was computed using the disc instability code of \citet{1998MNRAS.298.1048H}, assuming Solar composition.  The evolution of the system is then tracked in the HRD using the resulting  $G$ and \bprp\ lightcurves.

We chose two sets of DN binary parameters as representative of the typical lightcurves produced by the disc instability model. One below the period gap with $P_{\rm orb}=88\rm\,min$, $M_{1}=0.6\rm\,M_{\odot}$, $M_{2}\approx0.05\rm\,M_{\odot}$, $\alpha_{\rm c}=0.04$, and $\alpha_{\rm h}=0.2$. The other above the period gap with $P_{\rm orb}=6\rm\,hr$, $M_{1}=1.0\rm\,M_{\odot}$, $M_{2}\approx0.6\rm\,M_{\odot}$, $\alpha_{\rm c}=0.02$, and $\alpha_{\rm h}=0.1$. In both cases we assumed that the white dwarf has a dipole moment $\mu=2\times 10^{30}\rm\,g\,cm^{3}$, which truncates the inner disc in quiescence. For the short $P_{\rm orb}$ system we computed lightcurves for $\dot{M}_{t}=2.5$, 10, and $100\times 10^{15}\rm\,g\,s^{-1}$. For the long $P_{\rm orb}$ system, we computed lightcurves for $\dot{M}_{t}=1.5$, 15, and $150\times 10^{15}\rm\,g\,s^{-1}$. These values of $\dot{M}_{\rm t}$ roughly sample the unstable range between a cold stable disc and a hot stable disc. The lightcurves are shown in the appendix (Fig.~\ref{fig:lightc}). The parameters were chosen to be close to those used by \cite{2020A&A...636A...1H} to track the evolution of SS Cyg and VW Hyi in $B-V$ and $V$ using the same DIM code. \object{VW Hyi} is identified as a CV by \gaia (but is not in our sample). It stays around \bprp$\approx 0.5$ and $M_G\approx 10$, which is consistent with a quiescent state. \object{SS Cyg} is not identified as a variable by \gaia so we do not have access to its epoch photometry.

The tracks in Fig.~\ref{fig:loop} take into account all sources of light. Each track forms a hysteresis pattern along which the system evolves anticlockwise, as discussed in \citet{2020A&A...636A...1H}.  The location of the outburst peak is in the region of steady discs, as expected from the high $\dot{M}$ and steady-like $T_{\rm eff}$ radial distribution at the outburst peak. The location in quiescence is close to the star symbol representing the combined light from the stellar components, as expected from the reduced contribution of the disc. The decline from the outburst is bluer than in outburst, again as expected as the disc cools systematically by propagating inwards from the outside whereas outbursts mostly start inside and propagate outwards. 

Each track has 100 dots that uniformly sample the cycle lightcurve in time. They give a measure of the probability of finding the CV along its HRD track. Low $\dot{M}_{t}$ systems spend most of their time in quiescence so the dots concentrate on the quiescent portion of the track. High $\dot{M}_{t}$ systems close to the stability line are in outburst for a large fraction of their eruption cycle, so the dots are more spread out along the track in the HRD. Figure \ref{fig:HRDvariexamples} in the Appendix shows the tracks followed by two DN captured by \gaia DR3, one above and one below the period gap. Their overall behaviour matches well with the expectations from the model (Fig.~\ref{fig:loop}).

\section{Discussion}

\subsection{The hook}
    
We find that the \gaia HRD for stars with low extinction shows an intriguing feature (the `hook') at $M_G\sim4$ and $\bprp\sim0.2$ going from the main sequence to the hot subdwarf clump direction (`upper hook') and continuing parallel to the main sequence down to $M_G\approx 6$ (`lower hook'). We argue that the hook is mostly composed of CVs, with some contamination from sdBs on the upper hook (Fig.~\ref{fig:hrds}). Comparing to the \citet{Ritter03} catalogue shows that these are novae-like CVs. This is also supported by the location of NL CVs in Fig.~2 of \citet{2020MNRAS.492L..40A} and in Fig.~3 of \citet{2022ApJ...938...46A}. The spread in the location of NL CVs is somewhat larger in their HRD than with our selection. This (mostly redward) spread is very likely due to them using different, looser criteria on distance uncertainty, photometric quality and/or extinction. The yellow stars in the middle right panel of Fig.~\ref{fig:hrds}, illustrate the effect of using wider selection criteria.

Novae-likes are accreting at a high rate, so their disc is hot enough for hydrogen to be fully ionised and is thus stable against the thermal-viscous instability that triggers dwarf novae outbursts. The stability of a stationary disc depends on its size, since the temperature decreases with radius, and on the mass accretion rate through the disc since $T^{4}\propto\dot{M}$. We find that the stability region is relatively limited in the HRD when the disc size and binary parameters are constrained by the observed distribution of CVs. This stability region overlays well with the hook (Fig.~\ref{fig:steady}). It also appears rather stable against assumptions on the companion magnitude and colour (compare Fig.~\ref{fig:steady} to Fig.~\ref{fig:knigge}). 

The upper hook roughly corresponds to the stable region predicted by the DIM. It is limited in $M_G$ because it is rapidly bounded at low luminosity by the disc instability and at high luminosity by the mass transfer rate from the companion, which becomes unrealistically large ($\dot{M}> 10^{18}\rm\,g\,s^{-1}$). In the colour direction, there is some uncertainty on how far redward the hook extends because the CVs then have increasingly longer $P_{\rm orb}$, with the companion having an increasingly important contribution to the total light of the system. However, we did not take into account that the companion may become evolved in CVs with $P_{\rm orb}\geq 6\rm\,hr$, which introduces an uncertainty. Blueward, the CVs have shorter orbital periods, evolutionary models predict that the secular mass transfer rate from their companion drops, and the stability region turns to fainter magnitudes in the HRD diagram. In other words, the lack of CVs left of $4<M_{G}<6$ and $-0.5<\bprp<0$ implies $\dot{M}<10^{17}\rm\,g\,s^{-1}$, which is expected for short $P_{\rm orb}$ systems.

The lower hook, which extends parallel to the main sequence, is more difficult to explain. We did not take into account the effects of inclination and limb darkening on the disc contribution. These will make CVs appear fainter and redder:  such a systematic trend in the HRD is detectable using CVs \citep[see][]{2018AJ....156..198H}. Indeed, all novae-like CVs on the lower hook with $M_G>5.3$ in Fig.~\ref{fig:hrds} have $\cos{i}<0.5$. Inclination effects may thus partly explain some of the extension at the faintest magnitudes. Another way to extend the stability to fainter systems is if the inner disc is truncated, in the sense that there is no contribution from the accretion flow inward of a certain truncation radius. This truncation radius can be the result of disruption by the white dwarf magnetic field or the result of the disc switching to another type of flow, for instance magnetic wind accretion \citep{2019A&A...626A.116S}. These effects are all the more important that the mass accretion rate is low, which is what is expected for low $P_{\rm orb}$ systems. Figure~\ref{fig:steady} shows that this extends the stability region along the lower hook, with the bluest systems having the shortest orbital periods. Intermediate polars, which have discs truncated by the WD magnetic field, do populate the HRD to fainter magnitudes than NL in the HRD of \citet{2020MNRAS.492L..40A} and \citet{2022ApJ...938...46A}. 

These results may be used to identify outliers and intringuing CVs based on their location in the HRD. For example, the NL HRD plots in \citet{2020MNRAS.492L..40A} and \citet{2022ApJ...938...46A} show outliers that are very far from the stability region. They may be misclassified; they may be due to contamination due to unresolved nearby stars;  they may harbour odd companions; they may also represent new subclasses. For instance, the NL with $M_{G}> 10$ plotted  in the middle right panel of Fig.~\ref{fig:hrds} could be examples of cold and stable systems. The DIM predicts such systems can exist for low mass accretion rates if the disc is significantly truncated. The right hand side panel of Fig.~\ref{fig:steady} shows that such systems must be located in a thin strip (blue lines) close to and connecting to the white dwarf sequence. In fact, all the faint ($M_G>8$) systems identified as NL in the \citet{Ritter03} catalogue (see the middle right panel, Fig.~\ref{fig:hrds}) are also subtyped as polars or intermediate polars, indicating that their disc is truncated.

\subsection{Variability in the HRD}
Individual CVs will move in the HRD as they fluctuate or undergo outbursts. For accretion discs that remain globally stable, variations in the mass accretion rate through the disc move the system along well-defined tracks (dashed lines in Fig.~\ref{fig:steady}), on a timescale set by the viscous timescale $\sim 0.5$ to 1\,day. The observed tracks in the HRD of the brightest NL are entirely consistent with this behaviour (Fig. \ref{fig:brightestnovae}). This offers a way to distinguish CVs from sdBs, whose variability is mostly horizontal (in colour) on the HRD. \gaia DR3 gives access only to the lightcurves for sources classified in \gaia variability classes i.e. we only have access to epoch photometry for the sources identified as CVs, SBDs and short-timescale variables (the three classes present on the hook). Looking at the 11 stars with short time-scale variations on the hook (Tab.~\ref{table:hookS}), only 1 has sdB-like variability (that is small $M_G$ variability and large \bprp\ one). The other 10 have CV-like variability, meaning correlated $M_G$ and \bprp\ variability (Fig.~\ref{fig:hookS}). Of these 10 sources, 5 are already identified as CVs in Simbad (\object{HQ And}, \object{2MASS J09253483+4349179}, \object{V1084 Her}, \object{V393 Hya }, \object{EC 21263-4452}) and 5 are currently labelled as hot subdwarf candidates in Simbad. This method can be used to classify other systems as candidate NL CVs once their epoch photometry becomes available in DR4.

Dwarf novae outbursts occur when the disc becomes unstable to the thermal-viscous instability. The CV then traces a complex track in the HRD as the disc temperature evolves as a function of radius and time. To zeroth order, the system at outburst peak is located close to the stability region of the HRD because the disc then has a high $\dot{M}$ and a temperature distribution close to stationary. In quiescence, the disc is faint and the system lies on the HRD line formed by adding the contributions from the white dwarf and its stellar companion. The heatscatter plot of variability clearly shows two corresponding clusters, one associated with the hook in the NL region and one at fainter luminosities (right panel of Fig.~\ref{fig:hrds}).

Interpreting these clusters requires knowledge of the probability of catching each CV at a particular location and general knowledge of the CV population. The amount of time the system spends at each location strongly depends on the outburst duty cycle, with the duty cycle increasing as the mass transfer rate from the companion decreases (Fig.~\ref{fig:loop}). Typically, DN below the gap, such as WZ Sge systems, have a low $\dot{M}_{\rm t}$ and a long duty cycle such that a random observation is likely to place them towards the faint end of the HRD. Inversely, DN above the period gap such as U Gem systems are likely to be brighter and located in the upper half of the HRD. Figure \ref{fig:loop} also indicates that systems below the period gap can move through a larger region of the HRD than systems above the period gap. This suggests that the brighter cluster in Fig.~\ref{fig:hrds} (right panel) is mostly associated with CVs above the period gap, and that the fainter cluster is mostly associated with systems below the period gap, which is indeed what is observed, see e.g. Fig.~1 of \cite{2022ApJ...938...46A}. Finally, we note that DN run through their outburst loops counterclockwise in Fig.~\ref{fig:loop}. This should be the case for most systems as the temperature always drops from the outside in during decline from outburst, and generally rises from the inside out (outside in rises to outburst are also possible for high $\dot{M}$ systems, \citealt{Lasota:2001th}).

\section{Conclusion}
We attribute to CVs an intriguing feature in the \gaia HRD for low-extinction systems (the `hook'). This feature arises as a consequence of the disc instability model: the hook corresponds to the region where stable systems are located in the HRD. Unstable systems, giving rise to DN outbursts, trace loops counterclockwise in the HRD. The general behaviour is consistent with the location of the various CV subtypes in the HRD. These results can be used as a basis to pinpoint interesting outliers in the HRD, either due to their location or their tracks. These outliers may signal new subtypes such as cold, stable CVs with truncated discs, or may challenge the disc instability model.

\begin{acknowledgements}
The authors acknowledge support from the Centre National d'Etudes Spatiales (CNES). This research has made use of the Spanish Virtual Observatory (https://svo.cab.inta-csic.es) project funded by MCIN/AEI/10.13039/501100011033/ through grant PID2020-112949GB-I00.
This work has made use of data from the European Space Agency (ESA) space mission \gaia (\url{https://www.cosmos.esa.int/gaia}), processed by the \gaia Data Processing and Analysis Consortium (DPAC). Funding for the DPAC is provided by national institutions, in particular the institutions participating in the \gaia MultiLateral Agreement. 
This research has made use of the SIMBAD database, operated at CDS, Strasbourg, France.
\end{acknowledgements}

\bibliographystyle{aa}
\bibliography{gaiacv}

\begin{appendix}

\section{\gaia magnitudes from model CV systems \label{sec:GaiaMags}}
This Appendix describes the assumptions used to compute the $G$ magnitude and \bprp\ colour of the model CV systems presented in this work. We take into account light from the accretion disc, the companion, the white dwarf, and from the bright spot where matter from the L1 point intersects the accretion disc.  We do not take into account light from the boundary layer where material from the accretion disc settles onto the white dwarf. We expect the boundary layer to emit mostly in FUV and X-rays, making a negligible contribution to the \gaia magnitudes. We also neglect the possible impact of irradiation of the accretion disk or the stellar companion by the FUV/X-ray light from the white dwarf and its boundary layer, which would depend on further assumptions on the irradiation geometry, flux and albedo.

\subsection{The accretion disc}
The accretion disc magnitudes are computed as described in Section 2.1 of \citet{2018A&A...617A..26D}. The disc has a radial temperature profile, and each ring is assumed to radiate like a blackbody. This disc blackbody emission is converted to \gaia $G$ and \bprp\ using the filter response from the third \gaia data release as implemented in the SVO Filter Profile Service \citep{2012ivoa.rept.1015R, 2020sea..confE.182R}. This also gives the zeropoints of the 3 bandpasses : $Z_G=3228.75$\,Jy, $Z_{G_{BP}}=3552.5$\,Jy and $Z_{G_{RP}}=2554.95$\,Jy. The magnitudes also depend on the disc inclination $i$ to the observer. We marginalise over this parameter by assuming that the normal vector of the disc is randomly distributed and averaging over $i$.

\subsection{The stellar components}
Given an orbital period $P_{\rm orb}$ and white dwarf mass $M_1$, the companion star mass $M_2$ and radius $R_2$ are derived by assuming it fills its Roche lobe. The Roche lobe radius as a function of $q=M_2/M_1$ is approximated by the formula in \citet{1983ApJ...268..368E}. The mass, radius and luminosity $L_2$ of the companion are assumed to be related by the usual approximate main sequence relationships: $L_2\propto M_2^{3.5}$, $R_2 \propto M_2^{0.8}$ (in Solar units) for $M_2<\rm 1\, M_\odot$ and $R_2 \propto M_2^{0.57}$ for $M_2>\rm 1\, M_\odot$. We set the white dwarf effective temperature $T_1=12\,500\rm\,K$, corresponding to \bprp$\approx 0$. Its radius $R_1$ is derived from the \citet{1972ApJ...175..417N} mass radius relationship for white dwarfs. The \gaia magnitudes for both stellar components are then computed from $R$ and $L$ (or $T$) by assuming blackbody emisssion.

\begin{figure}
    \centering
    \includegraphics[width=0.33\textwidth]{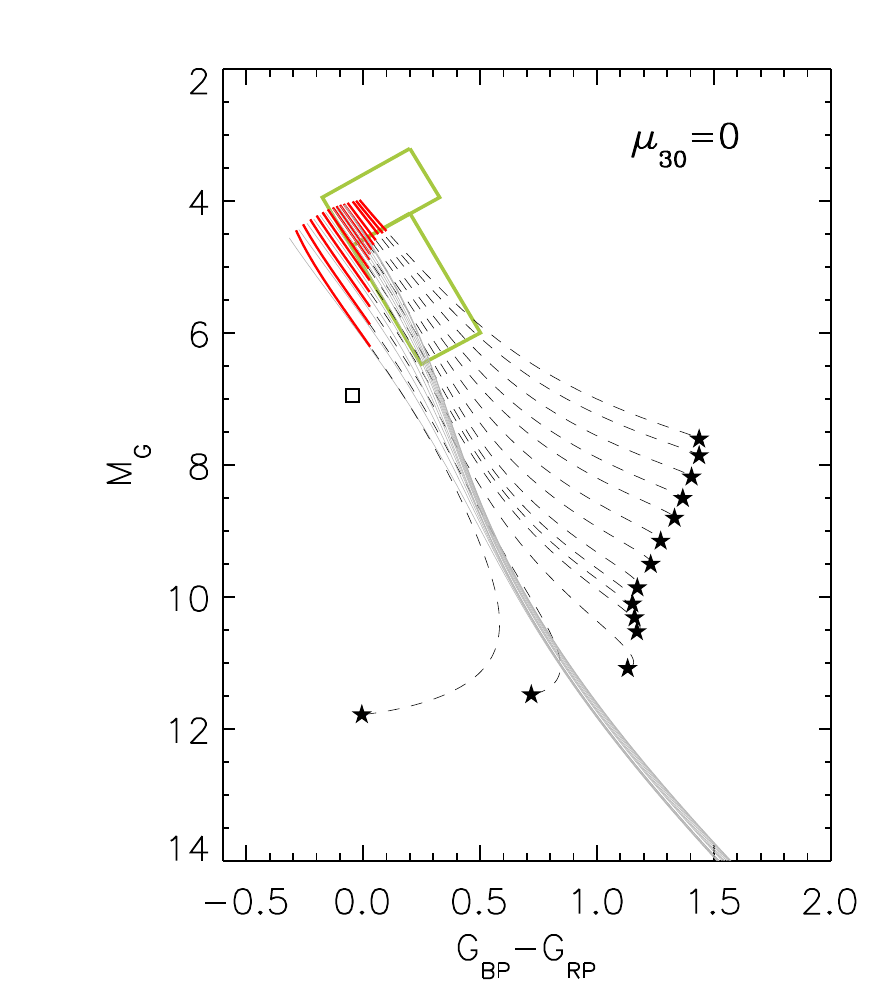}
     \caption{Same as Fig.~\ref{fig:steady}, using the companion star and white dwarf properties tabulated in \citet{2011ApJS..194...28K}. Here, $P_{\rm orb}$ sequences are plotted up to 5.6\,hr (instead of 7\,hr) for consistency with the tables.}
    \label{fig:knigge}
\end{figure}

This is rather simplistic, but works reasonably well. Indeed, we show in Fig.~\ref{fig:knigge} the results obtained by using the donor and white dwarf sequences derived specifically for CVs by \citet{2011ApJS..194...28K}. These are valid from the period minimum at 80\,min up to $P_{\rm orb}=5.6\rm\,hr$. In this case, we derived the \gaia magnitudes of the companion using the effective temperature \bprp\ colour relations and luminosity $M_G$ relations of \cite{Baraffe15}\footnote{\url{http://perso.ens-lyon.fr/isabelle.baraffe/BHAC15dir/}} for the main-sequence stars and of \cite{2006AJ....132.1221H}\footnote{\url{https://www.astro.umontreal.ca/~bergeron/CoolingModels/}} for white dwarfs. Comparing Fig.~\ref{fig:knigge} with Fig.~\ref{fig:steady} shows the same conclusions apply on the location of the stable zone (red lines in the HRD). 

Here, the accretion disc magnitudes do not converge to a single line at low $\dot{M}$ since $R_1$ varies with $P_{\rm orb}$ along the sequence. The jump in magnitude around $\rm M_G\approx 10$ is related to the period gap between $P_{\rm orb}=2$\,hr and 3\,hr.

\subsection{The bright spot}
The luminosity of the bright spot is given by the rate at which gravitational energy from the infalling L1 material must be dissipated:
\begin{equation}
L_{\rm BS} = \frac{GM\dot{M}_{\rm t}}{2}\left[\frac{1}{R_{\rm disc}}-\frac{1}{R_{\rm L1}}\right]
\end{equation}
where $R_{\rm disc}$ is the disc outer radius and $R_{\rm L1}$ is the distance to the white dwarf of the L1 Lagrange point. $\dot{M}_{\rm t}$ is the mass transfer rate from the companion throught the L1 point. We set the bright spot effective temperature to $T_{\rm BS}=13\,000\rm\,K$ (e.g. \citealt{2010MNRAS.402.1824C}). The \gaia magnitudes are calculated from $L_{\rm BS}$ and $T_{\rm BS}$ assuming blackbody emission. 

\section{Additional figures}

\begin{figure*}
    \centering
    \includegraphics[width=\columnwidth]{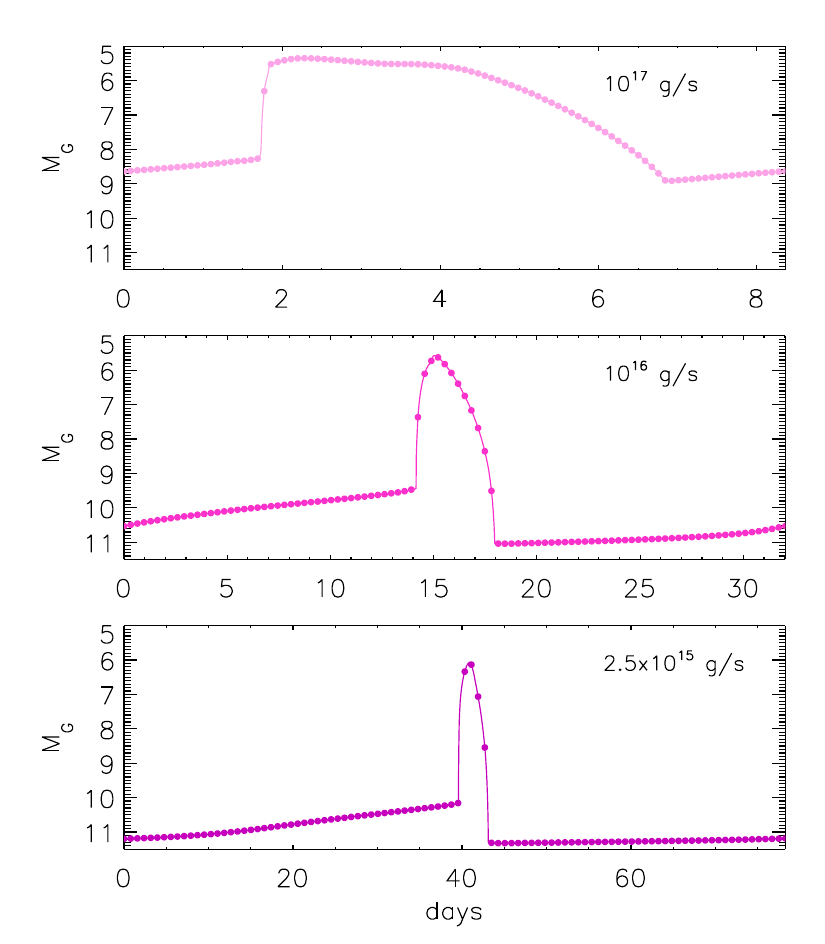}
    \includegraphics[width=\columnwidth]{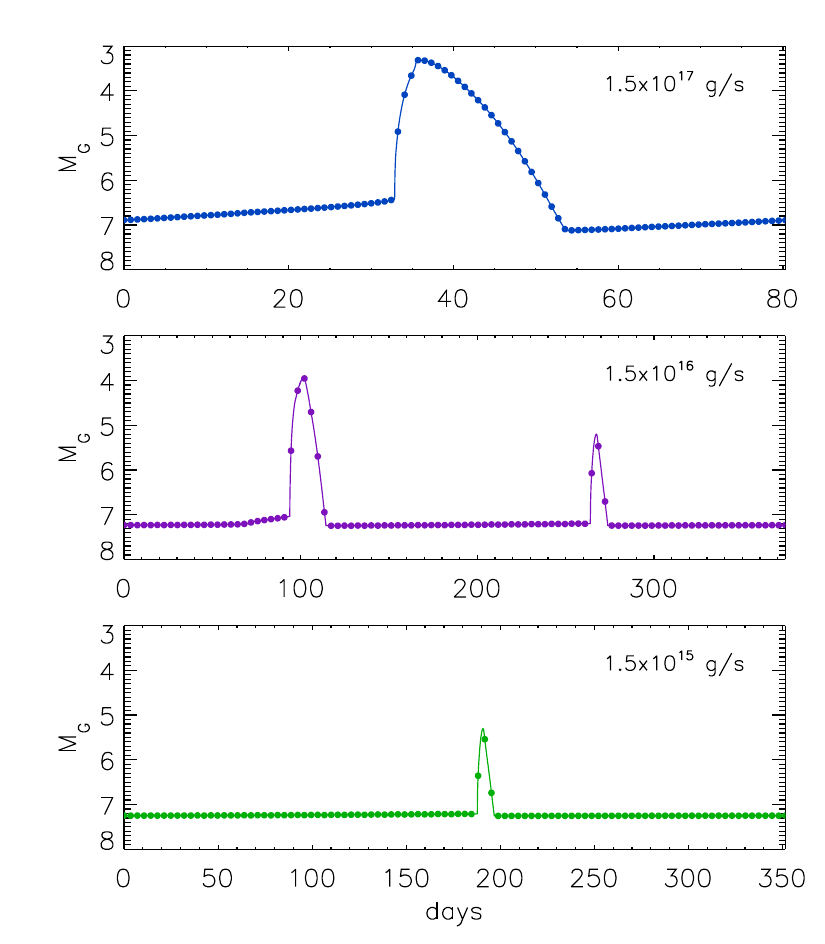}
     \caption{Lightcurves for the dwarf novae colour-magnitude tracks plotted in Fig.~\ref{fig:loop}. The recurrence timescale of the outburst cycle sets the time span of each panel. Left panels : lightcurves for the system with $P_{\rm orb}=88\rm\,min$. Right panels: lightcurves for the system with $P_{\rm orb}=6\rm\,hr$. The dots regularly sample in time the outburst cycle.}
    \label{fig:lightc}
\end{figure*}

\begin{figure*}
    \includegraphics[width=\textwidth]{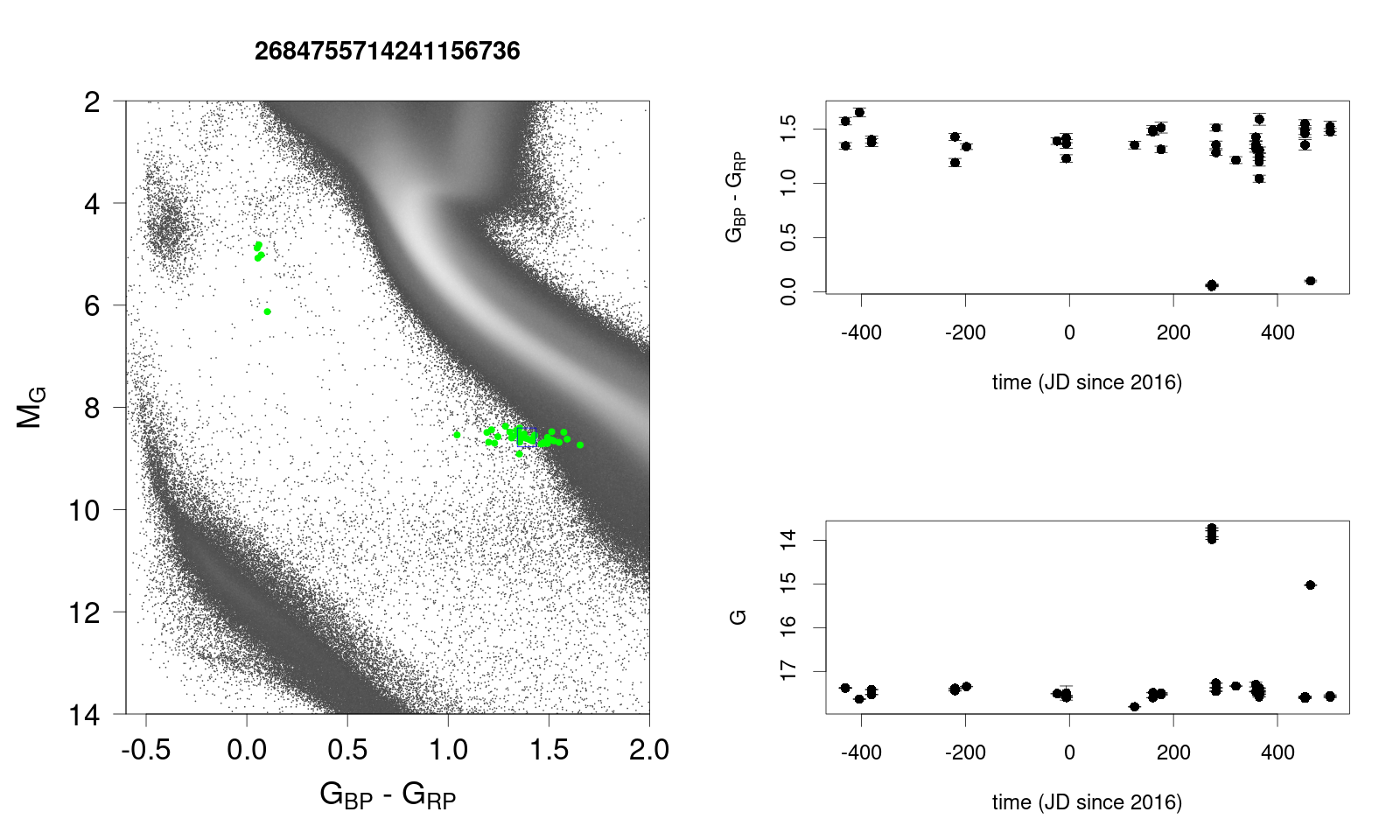}
    \includegraphics[width=\textwidth]{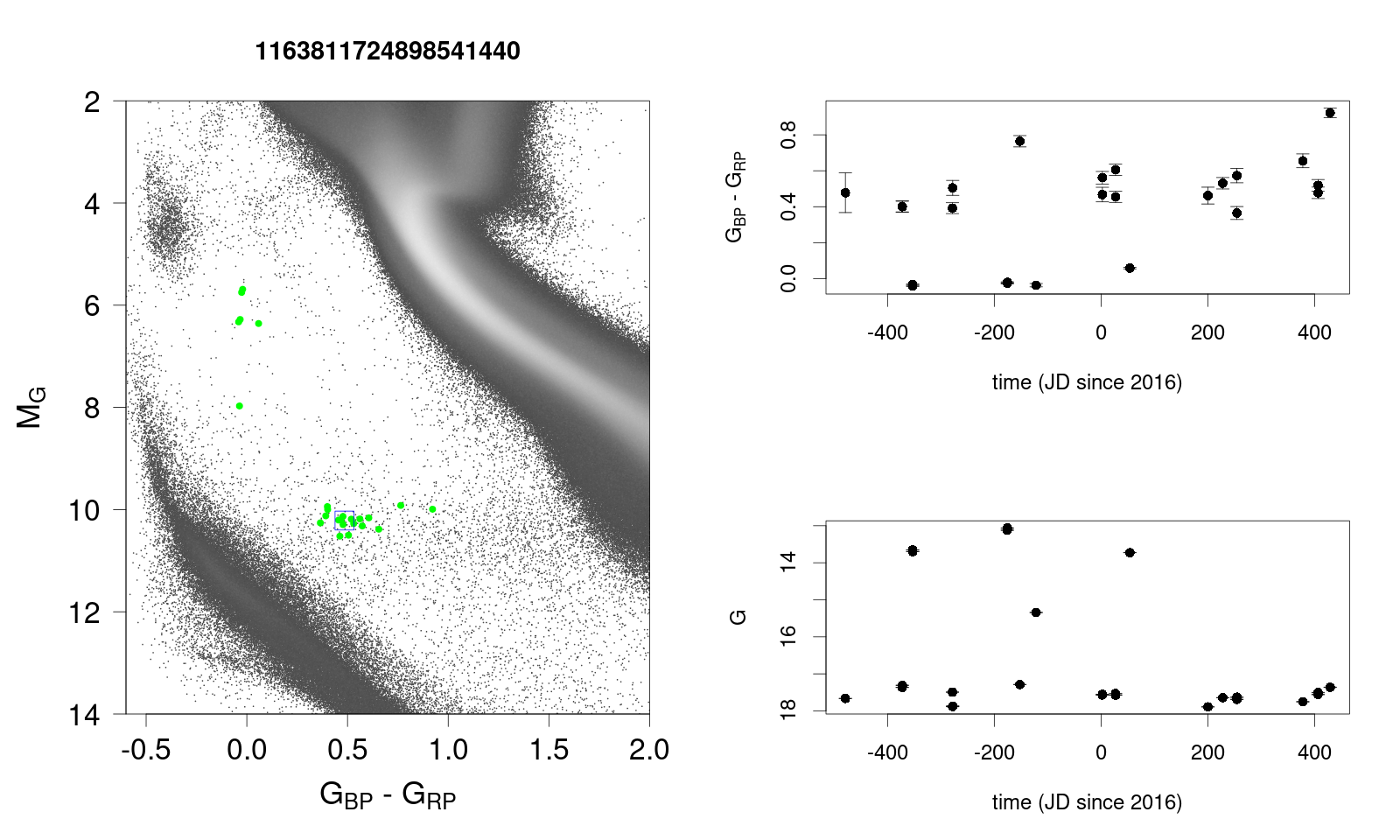}
    \caption{Two examples of Dwarf Novae captured by \gaia DR3.
    The green points indicate the positions in the HRD of each observation while the blue square indicates the mean photometry. 
    Top: \object{VZ Aqr} = \gaia DR3 2684755714241156736 with orbital period of 3.8h. 
    Bottom: \object{QW Ser} = \gaia DR3 1163811724898541440 with orbital period of 1.79h.
    }
    \label{fig:HRDvariexamples}
\end{figure*}

\begin{figure*}
    \includegraphics[width=0.9\textwidth]{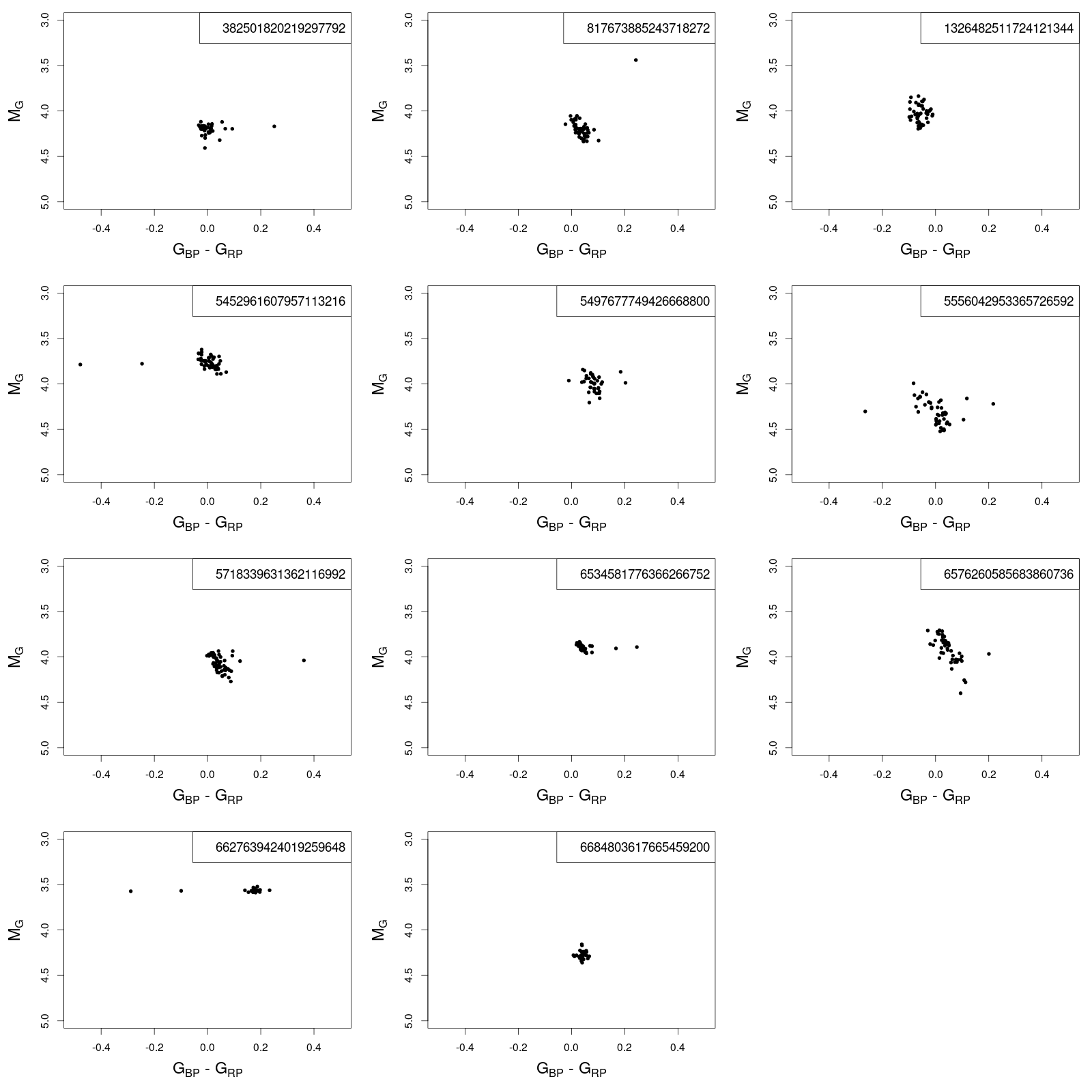}
    \caption{
    Variation in the HRD of upper hook stars with epoch photometry thanks to their short time scale variability detection in \gaia DR3.
    }
    \label{fig:hookS}
\end{figure*}

This Appendix contains information to complement the main text. Figure~\ref{fig:lightc} shows the lightcurves for the CV tracks plotted on the HRD in Fig.~\ref{fig:loop}. Figure~\ref{fig:HRDvariexamples} shows examples of the tracks left on the HRD by two dwarf novae, one above the period gap and one below. Table~\ref{table:hookS} and Fig.~\ref{fig:hookS} present the 11 variables in the upper-hook area that have epoch photometry available within \gaia DR3 thanks to their short time scale variability detection. One (6627639424019259648) has sdB like variability while the others have CV-like HRD variability with a clear correlation between $M_G$ and \bprp\ for the not already known CV. Only the variability profile of \gaia DR3 6684803617665459200 is unclear.

\begin{table}
\begin{tabular}{rcc}
\hline
source\_id & VarG & Variability \\
\hline
\object{Gaia DR3 382501820219297792}&1.0&known CV\\ % HQ And
\object{Gaia DR3 817673885243718272}&2.0&known CV\\ %2MASS J09253483+4349179
\object{Gaia DR3 1326482511724121344}&1.4&known CV\\ %V1084 Her
\object{Gaia DR3 5452961607957113216}&1.0&known CV\\ %V393 Hya 
\object{Gaia DR3 5497677749426668800}&1.6&CV like\\ %Hot Subdwarf Candidate
\object{Gaia DR3 5556042953365726592}&1.5&CV like\\ %Hot Subdwarf Candidate
\object{Gaia DR3 5718339631362116992}&1.7&CV like\\ %Variable Star
\object{Gaia DR3 534581776366266752}&1.4&CV like\\ %Hot Subdwarf Candidate
\object{Gaia DR3 6576260585683860736}&1.4&known CV\\ %EC 21263-4452
\object{Gaia DR3 6627639424019259648}&0.5&sdB like\\ % Hot Subdwarf
\object{Gaia DR3 6684803617665459200}&1.5&?\\ %Hot Subdwarf Candidate
\hline
\end{tabular}
\caption{Stars in the upper hook with epoch photometry thanks to their short time scale variability detection in \gaia DR3.}
\label{table:hookS}
\end{table}

\end{appendix}

\end{document}